\documentclass[12pt]{article}

\usepackage{amssymb}
\usepackage{amsmath}
\usepackage{latexsym}
\usepackage{yfonts}

\catcode`@=11
\renewcommand{\section}{\@startsection{section}{1}{0pt}{\medskipamount}
{\medskipamount}{\large\bf}} \numberwithin{equation}{section}
\catcode`@=12
\def\beq{\begin{eqnarray}}    %%%  begequation/eqnarray
\def\eeq{\end{eqnarray}}      %%%  endequation/eqnarray

\def\ln{\,\mbox{ln}\,}                  %%% log
\def\det{\,\mbox{det}\,}                %%% determinant
\def\pa{\partial}                       %%% partial
\def\={\ =\ }

\begin{document}
\begin{center}

{\Large\bf Generalized sigma model with dynamical antisymplectic
potential and non-Abelian de Rham's differential}

\vspace{18mm}

{\large Igor A. Batalin$^{(a,b)}\footnote{E-mail:
batalin@lpi.ru}$\;, Peter M. Lavrov$^{(b, c)}\footnote{E-mail:
lavrov@tspu.edu.ru}$\; }

\vspace{8mm}

\noindent ${{}^{(a)}}$
{\em P.N. Lebedev Physics Institute,\\
Leninsky Prospect \ 53, 119991 Moscow, Russia}

\noindent  ${{}^{(b)}} ${\em
Tomsk State Pedagogical University,\\
Kievskaya St.\ 60, 634061 Tomsk, Russia}

\noindent  ${{}^{(c)}} ${\em
National Research Tomsk State  University,\\
Lenin Av.\ 36, 634050 Tomsk, Russia}

\vspace{20mm}

\begin{abstract}
\noindent
For topological sigma models, we propose that their local Lagragian density is allowed to depend
non-linearly on the de Rham's "velocities" $D Z^{A}$. Then, by differentiating
the Lagrangian density
with respect to the latter de Rham's "velocities", we define a "dynamical"
anti-symplectic potential,
in terms of which a "dynamical" anti-symplectic metric is defined, as well.
We define the local and the
functional antibracket via the dynamical anti-symplectic metric.
Finally, we show that the generalized
action of the sigma model satisfies the functional master equation, as required.
\end{abstract}

\end{center}

\vfill

\\

\newpage

\section{Introduction}
When formulating a topological sigma model, one proceeds usually with an anti-symplectic
configuration space $\{Z^{A} | \varepsilon(  Z^{A} ) =: \varepsilon_{A}  \}$
whose anti-symplectic
potential $V_{A}( Z ),  \varepsilon( V_{A} )  =: \varepsilon_{A} + 1$,
is given originally \cite{AKSZ,SS,H,CF,LO,BM,BM5,BLN,BG,AG,BL}. At the same
time, it is assumed usually that the kinetic part of the original local
Lagrangian density $\mathcal{ L }$
has the form $ - ( D Z^{A} )  V_{A}( Z )$, where $D$ is the de Rham's differential.

In the present paper, we generalize  the local original Lagrangian density as to take the form
$\mathcal{ L }( Z, D Z )$ allowed to depend non-linearly on the de Rham's "velocities" $D Z^{A}$.
Then, we define a "dynamical" anti-symplectic potential, $V_{A}( Z, D Z )$
as the derivatives of the
new Lagangian density with respect to the mentioned de Rham's "velocities". We define a local
anti-symplectic metric in its covariant components $E _{AB}( Z, D Z )$ as
the standard vorticity of
the "dynamical" anti-symplectic potential $V_{A}( Z, D Z )$,
in terms of explicit $Z^{A}$ – derivatives.

We define both the local and the functional antibracket by the standard formulae via the
"dynamical" anti-symplectic metric in its contravariant components $E^{AB}( Z, D Z )$. Finally,
we show that the new action $\Sigma  =:  \int d\mu \mathcal{ L }( Z, D Z )$
satisfies the functional
master equation, provided the function
$S( Z, D Z )  =: \mathcal{ L }( Z, D Z ) +  DZ^{A} V_{A}( Z, D Z )$
satisfies the local master equation.

\section{Non-Abelian de Rham's differential}

Let $\Gamma$ be an intrinsic configuration super-manifold,
\beq
\label{GSa1}
\Gamma  =:  \{ X^{a}, C^{a} | \varepsilon( X^{a} ) = 0,\;
 \varepsilon( C^{a} ) = 1,  a = 1, . . . , 2 m \}.    %   (1.1)
\eeq
Let $D$ be a non-Abelian de Rham's differential, as defined by the conditions
\beq
\label{GSa2}
\varepsilon( D )  =  1, \quad  D^{2}  =  \frac{1}{2}  [ D, D ]  =  0,
\quad D  = -  D^{ \dagger },    %     (2.1)
\eeq
whose solution is sought  for in the form,
\beq
\label{GSa3}
D=:  C^{a} \Lambda_{a}^{b}( X ) \frac{ \partial }{ \partial X^{ b} }
+
  \frac{1}{2}  C^{b} C^{a} \mathcal{ U }_{ab}^{\;\!d} ( X )
\frac{ \partial }{ \partial C^{d} },   %    (3.1)
\eeq
with $\Lambda_{a}^{b}$ being invertible, and $U_{ab}^{d}$ being antisymmetric
in its subscripts $a, b$ \cite{BM}.
The conditions (\ref{GSa2}) imply
\beq
\label{GSa4}
\Lambda_{a}^{c} \partial_{c} \Lambda_{b}^{d}  -
( a \;\leftrightarrow\; b )  =    \mathcal{ U }_{ab}^{\;\!c} \Lambda_{c}^{d},  %       (4.1)
\eeq
\beq
\label{GSa5}
( - \Lambda_{a}^{e}\partial_{e} \mathcal{ U }_{bc}^{\;\!d}  +
\mathcal{ U }_{ab}^{\;\! e} \mathcal{ U }_{ec}^{\;\!d}  )  +
{\rm cyclic\; perm}. ( a, b, c )  =  0.   %    ( 5.1)
\eeq
The Jacobi relation (\ref{GSa5}) provides for the integrability of
the Maurer-Cartan equation (\ref{GSa4}).
In terms of the Boson integration measure,
\beq
\label{GSa6}
d\mu( \Gamma )  =: \rho( X ) [dX] [dC], \quad  \rho  =:  \det( \Lambda^{-1} )  =
(  \det( \Lambda )  )^{-1},    % (6.1)
\eeq
the anti-Hermiticity of the differential $D$ implies
\beq
\label{GSa7}
\rho^{-1} \partial_{b} ( \rho \Lambda_{a}^{b} )  +  \mathcal{ U }_{ab}^{\;\!b}=0 .   % (7.1)
\eeq
In turn, it follows from (\ref{GSa7}),
\beq
\label{GSa8}
\int d\mu( \Gamma )  D F(\Gamma )  =  0,   \quad
F(\Gamma) \Big|_{ \partial \Gamma }  = 0 .    %   (8.1)
\eeq

\section{Generalized sigma model}

Let $\Gamma'$ be an antisymplectic phase space,
\beq
\label{GSb1}
\Gamma'  =:  \{ Z^{A} | \varepsilon( Z^{A} ) =
\varepsilon_{A}, \; A = 1, . . . , 2 N \} ,    %  (1.2)
\eeq
with $N$ being an equal number of Boson and Fermion variables among $Z^{A}$.
Let us define the local action of the generalized sigma model in the form,
\beq
\label{GSb2}
\Sigma  =:  \int  d\mu( \Gamma )  \mathcal{ L }( Z(\Gamma), D Z(\Gamma) ),   %   (2.2)
\eeq
with the measure $d\mu( \Gamma )$ being defined in (\ref{GSa6}).
A functional derivative of the action (\ref{GSb2}) has the form,
\beq
\label{GSb3}
\frac{ \delta \Sigma }{ \delta Z^{A}(\Gamma) }  =  \partial_{A} \mathcal{ L }  +
D V_{A}(-1)^{ \varepsilon_{A} }, \quad
V_{A}  =:  -  \frac{ \partial \mathcal{ L } }{ \partial ( D Z^{A} ) } ,   %    (3.2)
\eeq
in terms of explicit $Z^{A}$-derivatives $\partial_{A}$.
In turn, we have
\beq
\label{GSb4}
D V_{A} (-1)^{ \varepsilon_{A} } = D Z^{B} \partial_{B} V_{A} (-1)^{ \varepsilon_{A} } =
-  \partial_{ B }  V_{A}  D Z^{B}
(-1)^{ \varepsilon_{B}( \varepsilon_{A} + 1 ) }.     %   (4.2)
\eeq
By inserting that into the relation (\ref{GSb3}), we get
\beq
\label{GSb5}
\frac{ \delta \Sigma }{ \delta Z^{ A}(\Gamma) }  =
E_{AB}  D Z^{B}  (-1)^{ \varepsilon_{B} }  +  \partial_{A} S ,  %  (5.2)
\eeq
where
\beq
\label{GSb6}
&&\quad E_{AB}  =:  \partial_{A} V_{B}  -  \partial_{B} V_{A}
(-1)^{ \varepsilon_{A} \varepsilon_{B} }, \\ %    (6.2)
%\eeq
%\beq
\label{GSb7}
&&S  =:  \mathcal{ L }  -  V_{A} D Z^{A} (-1)^{ \varepsilon_{A} }=
(1-N_{DZ}) \mathcal{ L }, \\
\label{GSb8}
&&\qquad\qquad N_{DZ}=:DZ^A\frac{\pa}{\pa (DZ^A)}\;.    %  (7. 2)
\eeq

\section{Functional and local master equations}

Let the metric (\ref{GSb6}) be invertible, and $E^{AB}$ be its inverse,
\beq
\label{GSc1}
E_{AB} E^{BC}  = \delta_{A}^{C} .    %     (1.3)
\eeq
Let $F( Z, D Z ), G( Z, D Z )$ be two arbitrary local functions;
their local antibracket is defined as
\beq
\label{GSc2}
( F, G ) ( Z, D Z )  =:  F( Z, D Z ) \overleftarrow{\partial }_{A}
E^{AB}( Z, D Z )  \overrightarrow{\partial }_{B}   G( Z, D Z),  %    (2.3)
\eeq
in terms of explicit $Z^{A}$ - derivatives $\partial_{A}$.

In turn, let $F[ Z ], G[ Z ]$ be two arbitrary functionals;
their functional antibracket is defined as
\beq
\label{GSc3}
( F, G )' [ Z ]  =:  F[ Z ]  \int  d\mu( \Gamma )
\frac{ \overleftarrow{\delta } }{ \delta Z^{A} }
E^{AB}( Z, D Z )  \frac{ \overrightarrow{\delta } }{ \delta Z^{B} }  G[ Z ] .   %      (3.3)
\eeq
Due to the definition of the $V_{A}$, the second in (\ref{GSb3}), together with
the definition (\ref{GSb6}) of the $E_{AB}$,  each of the antibrackets, (\ref{GSc2}) and
(\ref{GSc3}), satisfies its polarized Jacobi identity,\!
\footnote{Here in (\ref{GSc5}), $\Gamma'$ and $\Gamma^{''}$ mean ($X',C'$) and
($X^{''}, C^{''}$) as to stand
for ($X, C$) in (\ref{GSa1}). Not to be confused with (\ref{GSb1}).}
\beq
\label{GSc4}
(  (  Z^{A},  Z^{B}  ),  Z^{C}  )
(-1)^{ ( \varepsilon_{A} + 1 ) ( \varepsilon_{C} + 1 ) }  +
{\rm cyclic\; perm}. ( A, B, C )   =  0,   %     (4.4)
\eeq
\beq
\label{GSc5}
(  (  Z^{A}( \Gamma ),  Z^{B}( \Gamma' )  )',  Z^{C}( \Gamma^{''} )  )'
(-1)^{ ( \varepsilon_{A} + 1) ( \varepsilon_{C} + 1 ) }  +
 {\rm cyclic\; perm}. (  A, \Gamma;  B, \Gamma';  C, \Gamma^{''} )  =  0.  %     (4.5)
\eeq

Now, we are in a position to show that the functional master equation
\beq
\label{GSc6}
\frac{1}{2}  ( \Sigma, \Sigma )'  =  0,  %     ( 4.3)
\eeq
is satisfied as for the action $\Sigma[ Z ]$ (\ref{GSc2}),
provided that the local  master equation
\beq
\label{GSc7}
\frac{1}{2}  ( S, S )  =  0,   %    (5.3)
\eeq
is satisfied as for the function $S( Z, D Z )$ (\ref{GSb7}).
Indeed, by inserting the functional derivative (\ref{GSb5})
into the left-hand side of the equation (\ref{GSc6}), we have
\beq
\label{GSc8}
\frac{1}{2} ( \Sigma, \Sigma )'  =
\int  d\mu( \Gamma )  \left(  \frac{1}{2}  ( S, S )   +
D \mathcal{ L }  \right)  =  0, %   ( 6.3)
\eeq
due to the local master equation (\ref{GSc7}), together with the boundary condition,
\beq
\label{GSc9}
\mathcal{ L } |_{ \partial \Gamma }  =  0.     %     (7.3)
\eeq

As a simple example of the Jacobi identity as for the functional
antibracket (\ref{GSc3}), consider only the first equality in the formula  (\ref{GSc8}),
before the use of the equation (\ref{GSc7}),
\beq
\label{GSc15}
( \Sigma, \Sigma )'  =  \int  d\mu( \Gamma ) \;\! ( S, S ),   %    (4.15)
\eeq
which is valid for any functional of the form (\ref{GSb2}),
with $S$ defined by (\ref{GSb7}). Then, we get the non-polarized
form of the functional Jacobi identity,
\beq
\label{GSc16} (  (
\Sigma, \Sigma )',  \Sigma  )'  =  \int  d\mu( \Gamma )\;\!
(S,S)\;\!\Big(\frac{\overleftarrow{\partial} }{\partial Z^{A}} +
 \frac{ \overleftarrow{ \partial } }{\partial( D Z^{A})} D \Big)
 \Big(DZ^A+(S,Z^A)\Big)(-1)^{ \varepsilon_{A} }=0.
% \Big(  ( ( S, S ), S )  +
%( S, S )  \frac{ \overleftarrow{ \partial } }{ \partial ( D Z^{A} ) }
%D  ( D Z^{A} )  (-1)^{ \varepsilon_{A} } \Big)  =  0,   %     (4.16)
\eeq
In fact, the second equality in eq.(\ref{GSc16}) holds due
to the polarized functional Jacobi
identity (\ref{GSc5}) derived in Appendix A. The integrand in eq.(\ref{GSc16})
demonstrates explicitly
the structure of the original terms characteristic for the non-polarized  functional
Jacobi identity. Among other formal  properties involved, the simplest one is the
nilpotency of the de Rham's differential $D$, the second in the conditions
(\ref{GSa2}), as well
as the appearance of the boundary terms of the form "$D ( {\rm anything} )$",
and the local non-polarized Jacobi identity,
\beq
\label{GSc17}
( ( S, S ), S )  =  0,    %    (4.17)
\eeq
together with its $( D Z )$ - dual identity,
\beq
\label{GSc17z}
( S, S ) \frac{\overleftarrow{
 \partial } }{ \partial ( D Z^{A} ) } D ( Z^{A}, S )=  0. %       (4.13)
\eeq
Finally, if we introduce the nilpotent functional odd Laplacian,
\beq
\label{GSc10}
\Delta'  =:  \frac{1}{2}  ( \rho'[ Z ] )^{-1}
\int  d\mu( \Gamma )  ( -1)^{ \varepsilon_{A} }
\frac{ \delta }{ \delta Z^{A} } \rho'[ Z ]  E^{AB}( Z, D Z )
\frac{ \delta }{ \delta Z^{B} } ,    %    (8.3)
\eeq
with $\rho'[ Z ]$ being a local functional measure,
\beq
\label{GSc11}
\ln \rho'[ Z ]  =:  \int d\mu( \Gamma )  {\cal M}( Z, D Z ),   %      (9.3)
\eeq
then one can consider formally the functional quantum  master equation
\beq
\label{GSc12}
\Delta'  \exp\left\{ \frac{i}{\hbar} W \right\}  =  0 \; \Leftrightarrow \;
\frac{1}{2} ( W, W )'  +  \frac{\hbar}{i} \Delta' W  =  0.     %  (4.12)
\eeq
It looks natural to seek for a solution to the quantum master action functional $W$
in the form of a power series expansion in $i \hbar$,
\beq
\label{GSc13}
W  =  \Sigma  +  i \hbar\;\! W_{1}  +   . . .    .    %     (4.13)
\eeq
Then, to the $i \hbar$ order, it follows from (\ref{GSc12}),
\beq
\label{GSc14}
( \Sigma, W_{1} )'  -  \Delta' \Sigma  =  0.   %  (4.14)
\eeq Here in (\ref{GSc14}), the second term contains ill-defined
functional factor, $\delta(\Gamma,\Gamma)$. It seems rather difficult
to make sense to these expressions in a consistent way.  Thus, the
status of the functional quantum master equation (\ref{GSc12})
remains an open question.  Perhaps, the most realistic idea is that
the mentioned ill-defined factor might be canceled from the
functional path integral over the field-antifield variables
$Z^A(\Gamma)$.

\section*{Acknowledgments}
\noindent  The work of I. A. Batalin is supported in part by the
RFBR grants 17-01-00429 and 17-02-00317. The work of P. M. Lavrov is
supported in part by  the RFBR grant 15-02-03594.

\appendix
\section*{Appendix A}
%\section{}
\setcounter{section}{1}
\renewcommand{\theequation}{\thesection.\arabic{equation}}
\setcounter{equation}{0}

In this Appendix, we derive in detail the polarized Jacobi identity
(\ref{GSc5}) as for the functional antibracket. First, we rewrite
the relation (\ref{GSc5}) in the form\footnote{See footnote on page 4.}
\beq
\label{GSA1} E_{AB}( Z(
\Gamma ), D Z( \Gamma ) )  \delta( \Gamma, \Gamma' ) \frac{
\overleftarrow{\delta } }{ \delta Z^{C}( \Gamma^{''} ) } ( -1)^{
\varepsilon_{ A} \varepsilon_{C} }  +
{\rm cyclic\; perm}. ( A, \Gamma';  B, \Gamma;  C, \Gamma^{''} )  =  0.   %      (A.1)
\eeq
Now, the following relation holds
\beq
\nonumber
%\label{GSA2}
&&\frac{ \delta }{ \delta Z^{A}( \Gamma' ) }  V_{B}( Z( \Gamma ), D Z( \Gamma) )  -
( A, \Gamma'  \leftrightarrow B, \Gamma )  (-1)^{ \varepsilon_{A} \varepsilon_{B} }  =\\
\label{GSA2}
&&=\delta( \Gamma, \Gamma' )  E_{AB}( Z( \Gamma ), D Z( \Gamma) )  +\\
\nonumber
&&+(-1)^{ \varepsilon_{A} }
( D \delta( \Gamma, \Gamma' ) )
\left(  \frac{ \partial }{ \partial ( D Z^{A}( \Gamma ) ) }
V_{B}( Z( \Gamma ), D Z( \Gamma ) )  -
( A \leftrightarrow B )
( -1)^{ ( \varepsilon_{A} + 1 )( \varepsilon_{B} + 1 ) }\right)\!. %(A.2)
\eeq
The second term in the right-hand side in (\ref{GSA2}) equals to zero due
to the second in (\ref{GSb3}). Thus, we get from (\ref{GSA2}),
\beq
\nonumber
&&V_{B}( Z( \Gamma ), D Z( \Gamma ) )
\frac{ \overleftarrow{\delta } }{ \delta Z^{A}( \Gamma' )}
( -1)^{ \varepsilon_{A} \varepsilon_{B} }  -
V_{A}( Z( \Gamma' ), D' Z( \Gamma' ) )
\frac{\overleftarrow{ \delta } }{ \delta Z^{B}( \Gamma ) }  =\\
\label{GSA3}
&&=\delta( \Gamma, \Gamma' )  E_{AB}( Z( \Gamma ), D Z( \Gamma ) ).   %  (A.3)
\eeq By expressing the right-hand side from (\ref{GSA3}) and then
inserting into the left-hand side in (\ref{GSA1}),  we observe that
each term cancels its counterpart in the cyclic sum, so that the
polarized Jacobi identity (\ref{GSc5}) is confirmed.

 Notice that the nilpotency of the functional odd Laplacian (\ref{GSc10})
implies the polarized Jacobi identities (\ref{GSc5}) to hold, together with
their consequence that the $\Delta'$ is a differentiation
as for the functional polarized antibracket,
\beq
\label{GSA4}
\Delta'  ( Z^{A}( \Gamma ),  Z^{B}(\Gamma' ) )'  =
( \Delta' Z^{A}( \Gamma ),  Z^{B}( \Gamma' ))'  -
( Z^{A}( \Gamma ),  \Delta' Z^{B}( \Gamma' ))'  (-1)^{ \varepsilon_{A} },   %  (A.4)
\eeq
and the property \cite{BBG},
\beq
\label{GSA5}
(\Delta'_{1})^2+{\rm ad}'\Big(\frac{1}{\sqrt{\rho'}}\;\!\Delta'_{1}\sqrt{\rho'}\Big)=0,
\quad
\Delta'_{1}=:\Delta' \big|_{\rho'=1}.
\eeq

\appendix
\section*{Appendix B}
%\section{}
\setcounter{section}{2}
\renewcommand{\theequation}{\thesection.\arabic{equation}}
\setcounter{equation}{0}

In this Appendix B, we consider in short the local measure,
\beq
\label{GSB1}
\Delta' \Sigma  =  \frac{1}{2}   \int
d\mu(\Gamma)\Big(\delta(\Gamma,\Gamma)
\partial_{A} ( Z^{A}, S ) (-1)^{\varepsilon_{A}}  +
\frac{\delta \ln \rho' }{\delta Z^{A}}\big(D Z^{A}  +
( S,Z^{A})\big)\Big).    %      (B.1)
\eeq
%As the numbers of Bosons and Fermions among $Z^{A}$ are equal to each
%other,
%\beq
%\label{GSB2}
%\delta_{A}^{A} (-1)^{ \varepsilon_{A} }  =  0.   %    (B.2)
%\eeq
In analogy with (\ref{GSb2}), (\ref{GSb5}), we have from (\ref{GSc11}),
\beq
\label{GSB3}
\frac{ \delta \ln \rho' }{ \delta Z^{A}( \Gamma ) }  &=&  \tilde{ E
}_{AB}  D Z^{B}  (-1)^{ \varepsilon_{B} }  +
\partial_{A}  \tilde{ S }, \\    %    (B.3)
%\eeq
%\beq
\label{GSB4}
\tilde{ E }_{AB}  &=:&  \partial_{A} \tilde{ V }_{B}   -
\partial_{B} \tilde{ V }_{A}  (-1)^{ \varepsilon_{A}
\varepsilon_{B} }, \\  %      (B.4)
%\eeq
%\beq
\label{GSB5}
\tilde{ V }_{A} & =:&  -  \frac{ \partial \mathcal{ M } }{ \partial (
D Z^{A} ) }, \\  %   (B.5)
%\eeq
%\beq
\label{GSB6}
\tilde{ S }  &=:&  ( 1  -  N_{ D Z } ) \mathcal{ M }.     %   (B.6)
\eeq
Due to the relation (\ref{GSB3}), the second term in (\ref{GSB1}) rewrites
in the form
\beq
\label{GSB7}
(S,\tilde{ S })+S\overleftarrow{\pa}_C\big(E^{CA}\tilde{E}_{AB}-\delta^C_B\big)DZ^B
(-1)^{\varepsilon_B}+D\big(2\tilde{V}_BDZ^B(-1)^{\varepsilon_B}+S+\tilde{S}\big),
\eeq
which is a natural counterpart to the formula (\ref{GSc8}).

\appendix
\section*{Appendix C}
%\section{}
\setcounter{section}{3}
\renewcommand{\theequation}{\thesection.\arabic{equation}}
\setcounter{equation}{0}

Here, we consider in short the Legendre transformation from the de Rham's
"velocities" $D Z^{A}$ to new antifield variables $Z^*_{A}$.
Provided that the Hessian matrix
\beq
\label{GSC1}
\frac{ \partial^{2}
\mathcal{ L }( Z, D Z ) }{ \partial ( D Z^{A} ) \partial ( D Z^{B} ) },   %    (C.1)
\eeq
is invertible, one can introduce new antifield variables $Z^*_{A}$, by resolving the
definition
\beq
\label{GSC2}
Z^*_{A}  =:  -  \frac{ \partial \mathcal{ L }( Z, D Z ) }{ \partial ( D Z^A ) },  %   (C.2)
\eeq
with respect to the de Rham's "velocities"" $( D Z^{A} )( Z, Z^*)$. Then, we consider
the function of $Z^{A}, Z^*_{A}$,
\beq
\label{GSC3}
S ( Z, Z^* )  =: \mathcal{ L }( Z, ( D Z )( Z, Z^*) ) +
( D Z^{A} )( Z, Z^* )\;\!  Z^*_{A},   %     (C.3)
\eeq
in terms of which the general equation of motion for any dynamical quantity $\mathcal{ O }
( Z, Z^*)$ reads
\beq
\label{GSC4}
D \mathcal{O }  +  ( S, \mathcal{ O } )_{ ext }  =  0,  %   (C.4)
\eeq
where the only nonzero elementary extended antibrackets are
\beq
\label{GSC5}
( Z^{A}, Z^*_{B} )_{ ext}  =:  \delta^{A}_{B}.     %     (C.5)
\eeq
As for the function (\ref{GSC3}) by  itself, we have
\beq
\label{GSC6}
D S  =  0,  \quad  ( S, S )_{ ext }  =  0. %  ( C. 6)
\eeq

By choosing in the general equation (\ref{GSC4}) $\mathcal{ O }  =  Z^{A}$, and then
$\mathcal{ O }  =  Z^*_{A}$, we have
\beq
\label{GSC7}
D Z^{A}  =  -  (  S,  Z^{A} )_{ ext },\quad
D Z^*_{A}  =  -  (  S,  Z^*_{A} )_{ ext } .    %    (C.7)
\eeq
These anticanonical equations are certainly generated by the usual action
linear in the de Rham's "velocities",
\beq
\label{GSC8}
\Sigma  =:  \int  d\mu( \Gamma )  \big(  Z^*_{A}  D Z^{A}  (-1)^{ \varepsilon_{A} }  +
S( Z, Z^*)  \big).    %   (C.8)
\eeq
Thus, provided  that the Hessian matrix (\ref{GSC1}) is invertible, we arrive at the usual
situation although formulated within the phase space whose dimension is
twice as more than the one of the original antisymplectic phase space (\ref{GSb1}).

\end{document}